\documentstyle[aps,prl,preprint,floats,epsfig]{revtex}
\newcommand{\Ktwost}{K_2^*(1430)}
\newcommand{\Ebeam}{E_{\mathrm{beam}}}
\newcommand{\coshel}{\cos\theta_{\mathrm{hel}}}
\newcommand{\thetahel}{\theta_{\mathrm{hel}}}
\newcommand{\fbi}{\mathrm{fb}^{-1}}
\newcommand{\Mbc}{M_{\mathrm{bc}}}
\newcommand{\bsgamma}{b \to s \gamma}

\newlength{\MINIPAGEWIDTH} \MINIPAGEWIDTH=.48\textwidth
\newcommand{\NBB}{\mbox{$22.8 \times 10^6$}}

\newcommand{\LumONRES}{21.3}

\newcommand{\YktwostgammN}{29.1 \pm 6.7 {\,}^{+2.4}_{-1.9}}
\newcommand{\YktwostgammNbsgamma}{0.4 \pm 0.3}

\newcommand{\YktwostgammNtrue}{20.1 \pm 10.5}
\newcommand{\Ykstpigamm}{46.4 \pm 7.3 {\,}^{+1.6}_{-2.7}}
\newcommand{\YkstpigammBKG}{5.8 \pm 2.2 {\,}^{+0.3}_{-0.8}}
\newcommand{\YkstpigammSUBTbsgamma}{0.9 \pm 0.6} % bsgamma
\newcommand{\YkstpigammSUBTSUBT}{39.7 \pm 7.4 {\,}^{+1.7}_{-2.6}}
\newcommand{\YkstpigammZ}{22.9 \pm 5.1 {\,}^{+1.0}_{-1.7}}
\newcommand{\Ykrhogamm}{24.5 \pm 6.4 {\,}^{+1.2}_{-2.3}}
\newcommand{\Ykrhogammbsgamma}{2.3 \pm 1.2}
\newcommand{\YkrhogammNSIGBOX}{4}
\newcommand{\YkrhogammNBKG}{1.19}
\newcommand{\EFFktwostgammN}{(6.99 \pm 0.55)\%}

\newcommand{\NktwostINkstpi}{2.6 \pm 1.4}
\newcommand{\EFFkstpigamm}{(3.13 \pm 0.47) \%}
\newcommand{\EFFkrhogamm}{(1.51 \pm 0.25) \%}
\newcommand{\EFFkoneAgamm}{(0.41 \pm 0.06) \%}

\newcommand{\BRULkstAgammN}{8.03 \times 10^{-5}}
\newcommand{\BRktwostgammNtrue}{%
  ( 1.26 \pm 0.66 \pm 0.10 ) \times 10^{-5}}
\newcommand{\BRkstpigamm}{%
  ( 5.6 \pm 1.1 \pm 0.9 ) \times 10^{-5}}
\newcommand{\ULkstpigammZ}{5.1 \times 10^{-5}}
\newcommand{\BRkrhogamm}{%
  ( 6.5 \pm 1.7 {\,}^{+1.1}_{-1.2} ) \times 10^{-5}}
\newcommand{\ULkoneAgammN}{9.6 \times 10^{-5}}
\begin{document}
%\doubles

\preprint{\tighten\vbox{\hbox{\hfil BELLE-CONF-0109}
}}

%\preprint{\vbox{ \hbox{ BELLE-CONF-01nn, Lepton Photon nn \hfill}}}
%@twocolumnfalse\endcsname

\title{\quad\\[1cm] \Large
 Observation of Radiative $B$ Meson Decays \\
 into Higher Kaonic Resonances}
\author{The Belle Collaboration}

\maketitle

% to make it single spaced
\tighten

\begin{abstract}
We have studied radiative $B$ meson decays
into higher kaonic resonances decaying into
a two-body or three-body final state,
using a data sample of $\LumONRES$ $\fbi$ recorded at
the $\Upsilon(4S)$ resonance with the Belle detector at KEKB.
For the two-body final state,
% we observe the
% $B \to \Ktwost \gamma$ signal for neutral and charged mode separately.
we extract the $B \to \Ktwost \gamma$ component
from an analysis of the helicity angle distribution,
and obtain
${\cal B}(B^0 \to \Ktwost^0 \gamma) = \BRktwostgammNtrue$.
%- For the three-body final state, for the first time
%- we observe
%- a $B \to K\pi\pi\gamma$ signal that is consistent
%- with a mixture of $B \to K^* \pi \gamma$ and $B \to K \rho \gamma$,
%- and determine their branching fractions to be
%- ${\cal B}(B \to K^* \pi \gamma; M_{K^*\pi}<2.0\mbox{~GeV}/c^2 )
%-  = \BRkstpigamm$
%- and
%- ${\cal B}(B \to K \rho \gamma; M_{K\rho} < 2.0\mbox{~GeV}/c^2 )
%-  = \BRkrhogamm$,
%- respectively.
For the three-body final state, 
we observe
a $B \to K\pi\pi\gamma$ signal that is consistent
with a mixture of $B \to K^* \pi \gamma$ and $B \to K \rho \gamma$.
This is the first time that $B \to K^* \pi \gamma$ and
$B \to K \rho \gamma$ have been observed separately.
We find their branching fractions to be
${\cal B}(B \to K^* \pi \gamma; M_{K^*\pi}<2.0\mbox{~GeV}/c^2 )
 = \BRkstpigamm$
and
${\cal B}(B \to K \rho \gamma; M_{K\rho} < 2.0\mbox{~GeV}/c^2 )
 = \BRkrhogamm$,
respectively.

\end{abstract}

{\renewcommand{\thefootnote}{\fnsymbol{footnote}}

%% >>>>>> LP2001 authorlist will go here
%% >>>>>> current version is /g8home/browder/lp2001/authorlist_lp2001.tex
%%% \input{authorlist_lp2001.tex}
% This is the author list for LP2001 and EPS conference papers.
% Added L.Y.Dong (IHEP), T.Hokuue (Nagoya), Y.Nakazawa (Nagoya).

\begin{center}
  K.~Abe$^{9}$,               % KEK
  K.~Abe$^{37}$,              % TohokuGakuin
  R.~Abe$^{27}$,              % Niigata
  I.~Adachi$^{9}$,            % KEK
  Byoung~Sup~Ahn$^{15}$,      % Korea
  H.~Aihara$^{39}$,           % Tokyo
  M.~Akatsu$^{20}$,           % Nagoya
  K.~Asai$^{21}$,             % Nara
  M.~Asai$^{10}$,             % Hiroshima
  Y.~Asano$^{44}$
% ...
\end{center}

\small
\begin{center}
% ...
$^{9}${High Energy Accelerator Research Organization (KEK), Tsukuba}\\
$^{10}${Hiroshima Institute of Technology, Hiroshima}\\
% ...
$^{13}${Institute for Theoretical and Experimental Physics, Moscow}\\
% ...
$^{15}${Korea University, Seoul}\\
% ...
$^{20}${Nagoya University, Nagoya}\\
$^{21}${Nara Women's University, Nara}\\
% ...
$^{27}${Niigata University, Niigata}\\
% ...
$^{37}${Tohoku Gakuin University, Tagajo}\\
% ...
$^{39}${University of Tokyo, Tokyo}\\
% ...
$^{44}${University of Tsukuba, Tsukuba}\\
% ...
\end{center}

\normalsize

%%%

\normalsize

\setcounter{footnote}{0}
\newpage

\normalsize

%%13.35.Dx, 14.60.Fg}
%\vskip2pc]                                                    % <---

%\onecolumn

\section{Introduction}

Radiative $B$ meson decay through the $\bsgamma$ process has been one of
the most sensitive probes of new physics beyond the Standard Model (SM).
The
inclusive picture of the $\bsgamma$ process is well established;
however, our knowledge of the exclusive final states in
radiative $B$ meson decays is
rather limited.  To date, we know that around 15\% of $\bsgamma$
can be accounted for by $B \to K^*(892)\gamma$ decays.
%process goes through $B \to K^*(892)\gamma$ decays.
In addition,
a relativistic form-factor
model calculation~\cite{PL-VeseliOlsson} predicts that another 20\% of
the $\bsgamma$ process should hadronize as
% go through
one of the seven known higher
kaonic resonances (Table~\ref{tab:resonances}).
CLEO has already reported an indication of the
$B \to \Ktwost \gamma$ signal ~\cite{PRL-CLEO-radb}.
Precision measurement of the inclusive $\bsgamma$
branching fraction  will require
%A detail study of $b \to s \gamma$ decay requires
detailed knowledge of such resonances, for example to model the decay
processes into multi-particle final states. In this
analysis, we study radiative $B$ meson decay processes into
higher kaonic resonances, which subsequently decay
into two-body or three-body final states.

We have analyzed a data sample that contains
$\NBB$ $B\bar{B}$ events.
The data sample corresponds to
an integrated luminosity of $\LumONRES$ $\fbi$ collected at
the $\Upsilon(4S)$ resonance with the Belle detector~\cite{NIM}
at the KEKB $e^+e^-$ collider~\cite{kekb}.
The beam energies are
%at the $\Upsilon(4S)$ resonance
$3.5\mbox{~GeV}$ for positrons and $8\mbox{~GeV}$ for electrons.

Belle is a general purpose detector with a typical laboratory polar angular
coverage between $17^\circ$ to $150^\circ$.
Charged tracks are reconstructed with
a 50 layer central drift chamber (CDC), and are then extrapolated and
refitted with a three layer double sided silicon vertex detector (SVD)
to provide precision track information for the decay vertex
reconstruction.  Particle identification, namely discrimination of kaons
from pions, is provided by combining information from silica aerogel
Cherenkov counters (ACC) and a time-of-flight counter system (TOF),
together with specific ionization ($dE/dx$) measurements from the CDC.
Photons are measured with an electromagnetic calorimeter (ECL) of 8736
CsI(Tl) crystals.  These detectors are surrounded by a 1.5 T
superconducting solenoid coil.

\begin{table}[bh]
 \begin{center}
  \catcode`;=\active \def;{\phantom{0}}
  \caption{Predicted branching fractions for
  radiative $B$ decays into kaonic resonances.}
  \begin{tabular}{cccl}
   & \multicolumn{2}{c}{Theoretical prediction $[\times 10^{-5}]$} & \\
   \multicolumn{1}{l}{mode}
   & \multicolumn{1}{l}{{\footnotesize Veseli-Olsson}\cite{PL-VeseliOlsson}} 
   & \multicolumn{1}{l}{{\footnotesize Ali-Ohl-Mannel}\cite{PL-AliOhl}}
   & sub-decay modes \\ \hline
    $B \to K^*(892)\gamma  $  & $;4.71 \pm 1.79$ & $;1.4-;4.9$ & $K\pi$ [$\sim$100\%] \\
    $B \to K_1(1270)\gamma  $  & $;1.20 \pm 0.44$ & $;1.8-;4.0$ & $K\rho$ [42\%], $K_0^*(1430)\pi$ [28\%], $K^*(892)\pi$ [16\%] \\
    $B \to K_1(1400)\gamma  $  & $;0.58 \pm 0.26$ & $;2.4-;5.2$ & $K^*(892)\pi$ [94\%], $K\rho$ [3\%] \\
    $B \to K^*(1410)\gamma  $  & $;1.14 \pm 0.18$ & $;2.9-;4.2$ & $K^*(892)\pi$ [$>$40\%], $K\pi$ [6.6\%] \\
    $B \to K_2^*(1430)\gamma$  & $;1.73 \pm 0.80$ & $;6.9-14.8$ & $K\pi$ [49.9\%], $K^*(892)\pi$ [24.7\%] \\
    $B \to K_2(1580)\gamma  $  & $;0.46 \pm 0.11$ & $;1.8-;2.6$ & $K^*(892)\pi$ \\
    $B \to K_1(1650)\gamma  $  & $;0.47 \pm 0.16$ & (not given) & $K \pi \pi$, $K \phi$ \\
    $B \to K^*(1680)\gamma  $  & $;0.15 \pm 0.04$ & $;0.4-;0.6$ & $K\pi$ [38.7\%], $K\rho$ [31.4\%], $K^*(892)\pi$ [29.9\%]
%    \mbox{total} & 17.6-36.4 & 10.44 \pm 3.78 \\ \hline
  \end{tabular}
  \label{tab:resonances}
 \end{center}
\end{table}

\section{Event Reconstruction}

We select events that contain a high energy (1.8 to 3.4 GeV in the
$\Upsilon(4S)$ rest frame) photon ($\gamma$) candidate inside the
acceptance of the barrel ECL
$(33^\circ<\theta_\gamma<128^\circ)$.  The photon candidate is required
to be consistent with an isolated electromagnetic shower, i.e., 95\% of
its energy is
% within a $3\times3$ cell around the local energy maximum and
concentrated in the central $3 \times 3$ crystals and
there is no associated charged track.  We combine it with other photon
clusters in the event and reject it if the invariant mass of the pair 
is consistent with a $\pi^0$ or $\eta$.

Kaonic resonance ($K_X$) candidates are formed
by combining one kaon with one or two pions,
in the two-body
$K^+\pi^-$, $K_S^0\pi^+$ and $K^+\pi^0$ final states
and in the three-body $K^+\pi^-\pi^+$ final state.
(Here and throughout the paper, charge conjugate modes are
implicitly included.)
%Every charged particle track of good quality is examined for the
%particle identification.
%A discrimination variable is defined as a
%likelihood ratio of the kaon and pion probabilities that are calculated
%for ACC, TOF and $dE/dx$ and then combined.
For every charged particle track of good quality, the
particle identification information is examined.
A likelihood ratio for the kaon and pion probabilities is calculated
by combining information from the ACC, TOF and $dE/dx$ systems.
We apply a tight cut with an
efficiency of 85\% for charged kaon candidates and a loose cut with an
efficiency of 97\% for charged pion candidates.  Neutral pion ($\pi^0$)
candidates are reconstructed from pairs of photons
that satisfy the following
requirements: the invariant two photon mass
is consistent with a $\pi^0$,
%within $\pm 16 \mbox{~MeV}/c^2$ of $M_{\pi^0}$,
each photon has more than 50 MeV energy,
the opening angle of two photons is less than
$40^{\circ}$ and
one of the photons should deposit its energy in more than one
crystal.  The $\pi^0$ momentum is recalculated with a $\pi^0$ mass constraint.
Neutral kaon ($K_S^0\to\pi^+\pi^-$)
candidates are reconstructed from two oppositely charged
tracks, whose invariant mass is consistent with $K_S^0$.
We require that
the $K_S^0$ candidate form a vertex displaced from the interaction point
and lie in a direction consistent with the $K_S^0$ momentum.

%Charged particle tracks of good quality are selected to be both
%charged pion and kaon candidates.  A kaon probability is formed from
%likelihood ratio variables that are individually calculated for ACC, TOF
%and CDC $dE/dx$.
%We require that the kaon probability is greater
%than 0.6 for kaon candidates and less than 0.9 for pion candidates.
%these retain 88\% of kaons and 96\% of pions.
%Neutral pion candidates
%are reconstructed from two photon clusters each with more than 50 MeV
%energy deposit, and required to be within $2\sigma$ of $M_{\pi^0}$.
%The $\pi^0$ momentum is refitted with a mass constraint.
%*******
%$K^0_S$ candidates
%are reconstructed from two charged tracks, whose invariant mass is
%within $3\sigma$ of $M_{K^0_S}$.  We require the $K^0_S$ candidate to
%form a displaced vertex from the interaction point and in a direction
%consistent with the $K^0_S$ momentum.
%For three-body decays, three charged tracks are required to form a
%vertex with the beam profile constraint.

We reconstruct $B$ meson candidates by forming two independent kinematic
variables: the beam constrained mass and the energy difference.  Both
variables are calculated in the $\Upsilon(4S)$ rest frame.  The beam
constrained mass is defined as
$\Mbc \equiv \sqrt{(\Ebeam)^2 - |\vec{p}_{K_X} + \vec{p}_\gamma|^2}$,
in which the photon energy is constrained to be
$E_\gamma = \Ebeam - E_{K_X}$.
This constraint improves the $\Mbc$ resolution by about 20\%, resulting
in an $\Mbc$ resolution of 3 $\mbox{MeV}/c^2$.
%The beam energy spread is the dominant contribution to the
%$\Mbc$ resolution.
The energy difference,
$\Delta E \equiv E_{K_X} + E_{\gamma} - \Ebeam$,
has an asymmetric resolution, mainly due to energy leakage from the
counters and energy loss in the inner material.  We apply a cut
of $-100\mbox{~MeV} <\Delta{E}<75\mbox{~MeV}$,
which is about a $2 \sigma$ cut on the higher side,
and which removes around 25\% of events on the lower side.

The largest background source is continuum light quark-pair
($q\bar{q}$) production,
in which the high energy photon mainly comes from the
initial state radiation
($e^+e^- \to q\bar{q}\gamma$)
and decays of neutral hadrons $(\pi^0, \eta, \ldots)$.
In order to reduce such background, we form a
Fisher discriminant which we call the {\it Super Fox-Wolfram} (SFW)
variable~\cite{PL-xsgamma},
\[
 {\it SFW} = \alpha_2 R_2^{\mathrm{major}} + \alpha_4 R_4^{\mathrm{major}}
 + \sum_{l=1}^4 \beta_l R_l^{\mathrm{minor}},
\]
\vspace*{-7mm}
\[
 R_l^{\mathrm{major}} = \frac{\sum_i |p_i||p_\gamma|P_l(\cos\theta_{i\gamma})}{ \sum_i |p_i||p_\gamma|}
  \mbox{,~~~}
 R_l^{\mathrm{minor}} = \frac{\sum_{i,j} |p_i||p_j|P_l(\cos\theta_{ij})}{
               \sum_{i,j} |p_i||p_j|},
\]
where $l$ runs from 1 to 4 for the Legendre function $P_l$ and
$i$, $j$ run over all the neutral and charged tracks that are not used
to form the $B$ candidate,
and coefficients $\alpha_i$ and $\beta_i$ are optimized to maximize
the discrimination.
The variable is calculated in the signal $B$
candidate rest frame
rather than the $\Upsilon(4S)$ rest frame in order to eliminate
any correlation with $\Mbc$.
% in which we find the correlation with $\Mbc$ is
%significantly reduced than in the $\Upsilon(4S)$ rest frame.
For further
background suppression, we form a likelihood ratio (LR) from the SFW
variable and the $B$ meson flight direction ($\cos\theta_B$),
\[
 {\mathit{LR}(SFW,\cos\theta_B)} =
  \frac{p^{\mathrm{sig}}(SFW)p^{\mathrm{sig}}(\cos\theta_B)}
  {p^{\mathrm{sig}}(SFW)p^{\mathrm{sig}}(\cos\theta_B)
  +p^{\mathrm{bg}}(SFW)p^{\mathrm{bg}}(\cos\theta_B)},
\]
where $p^{\mathrm{sig}}$ and $p^{\mathrm{bg}}$ are the
probability density functions (PDF) for signal and background.
The PDFs for the SFW are parametrized with an asymmetric Gaussian function.
%determined from a Monte Carlo (MC) simulation.
For $\cos\theta_B$, we use
$p^{\mathrm{sig}}(\cos\theta_B) \propto \sin^2\theta_B$ for signal and
a flat distribution for background.
We require $LR>0.7$ for the two-body $K_X$ final states and
$LR>0.9$ for the three-body $K_X$ final states.
The LR cut efficiency is determined from the
$B \to D\pi$ data samples using parametrizations
determined from the signal and continuum
$q\bar{q}$ Monte Carlo (MC) simulation samples.

The background from other $B$ meson decays is examined with
the corresponding MC samples.
We find a negligible contribution from hadronic charmless decays.
%Other $b \to s \gamma$ decays
Background from $b \to c$ decays
makes a non-negligible contribution in the sideband of negative
$\Delta E$ especially at high $K_X$ mass ($M_{K_X}>1.5 \mbox{~GeV}/c^2$),
but does not contribute in the signal region.
For the three-body final states,
there is a small contribution from $B \to K^*(892) \gamma$ decays
especially in the positive $\Delta E$ region;
this contribution is removed by rejecting candidates if
$\Mbc$ and $\Delta E$ calculated from $K^+\pi^-\gamma$
falls into the $K^*\gamma$ signal region.
%Other $b \to s \gamma$ backgrounds has a non-negligible
%contribution both as a combinatoric background
%source and also as a correlated background source,
%for which we assign the uncertainty as a systematic error.
Cross-feed from other $b \to s \gamma$ final states
is not negligible especially for the $K\pi\pi\gamma$ final state.
%We estimate it using MC sample and subtract it from the yield.
These contributions are estimated
using an inclusive $\bsgamma$ MC sample,
and subtracted from the signal yield.

We extract the signal yield from a fit to the $\Mbc$ distribution.
The shape is modeled as a sum of a Gaussian function for the signal
and a threshold-type function (ARGUS function~\cite{argus})
for the combinatoric background contribution.
The normalizations are floated  for both components.
The signal function is fixed from
the $B \to D \pi$ data.  The
background function is determined from $\Delta E$ sideband data
in the range $0.1\mbox{~GeV} < \Delta E < 0.5 \mbox{~GeV}$.
Using a continuum $q\bar{q}$ MC sample, we check that the
background shape has no visible correlation with $\Delta E$.  In order to
estimate the systematic error of the fitting procedure, we vary the mean
and width of the signal shape by $\pm 1 \sigma$ and use the background shape
from other sources, namely the continuum MC sample and
the LR sideband ($LR<0.3$) in which the signal
contribution is negligible.  We assign the
largest deviation as the systematic error of the signal yield.  As a
cross-check, we fit the $\Delta E$ distribution with a signal shape
from MC and a linear function for background
with a slope determined from the $\Mbc$ sideband
($5.2 \mbox{~GeV}/c^2 < \Mbc < 5.26 \mbox{~GeV}/c^2$),
and obtain consistent results.

\section{Analysis of $B \to \Ktwost \gamma$}

The $B \to \Ktwost \gamma$ analysis is performed by
%forming $\Ktwost$ from $K\pi$,
requiring the $K\pi$ invariant mass to be within
$\pm 125~\mbox{MeV}/c^2$ of the nominal $\Ktwost$ value.
The results of fits to the $\Mbc$ distributions
are shown in Fig.~\ref{fig:mbfit-k2st-data},
separately for the neutral and charged modes.
%The signal yields are $\YktwostgammN$ events
%for the neutral mode and $\YktwostgammC$ events for the
%charged mode.
The signal yield is $\YktwostgammN$ events for the neutral mode,
of which the contribution from other $\bsgamma$ decays is estimated
to be $\YktwostgammNbsgamma$ events; we also see some indication
of a signal in the charged mode.
These are consistent with the yields calculated
from the $\Delta E$ distributions shown in Fig.~\ref{fig:defit-k2st-data}.

\begin{figure}[!tb]
 \begin{center}
  \epsfxsize 0.45\textwidth \epsfbox{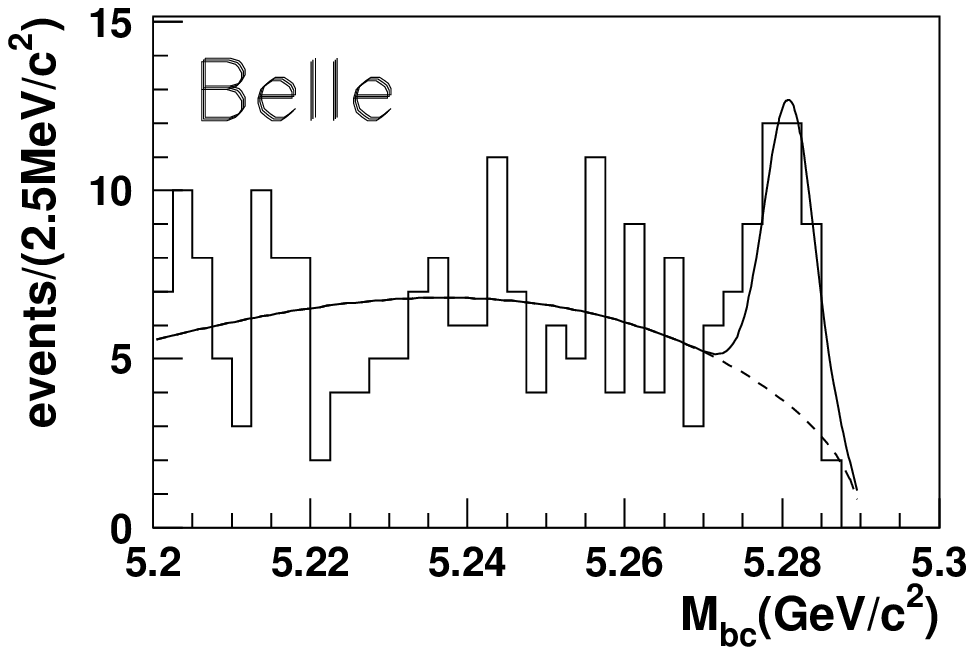}
  \hfil
  \epsfxsize 0.45\textwidth \epsfbox{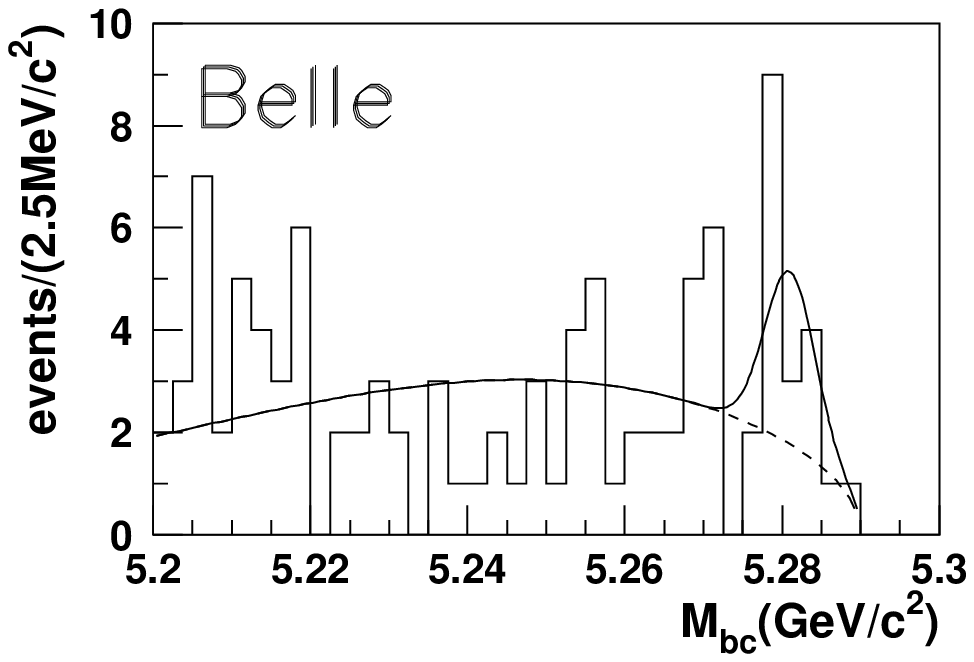}
  \caption{The $\Mbc$ distributions for $B^0 \to \Ktwost^0 \gamma$ (left)
  and $B^+ \to \Ktwost^+ \gamma$ (right) candidates.
  The solid line is the fitting result.
  The background component is shown as the dashed line.}
  \label{fig:mbfit-k2st-data}
 \end{center}
\end{figure}

\begin{figure}[!tb]
 \begin{center}
  \epsfxsize 0.45\textwidth \epsfbox{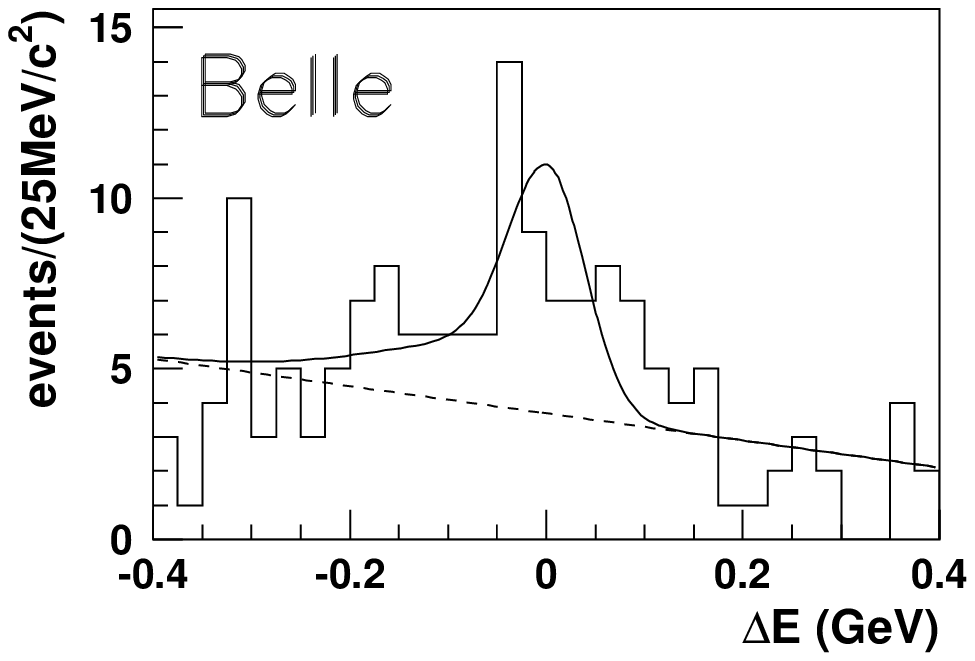}
  \hfil
  \epsfxsize 0.45\textwidth \epsfbox{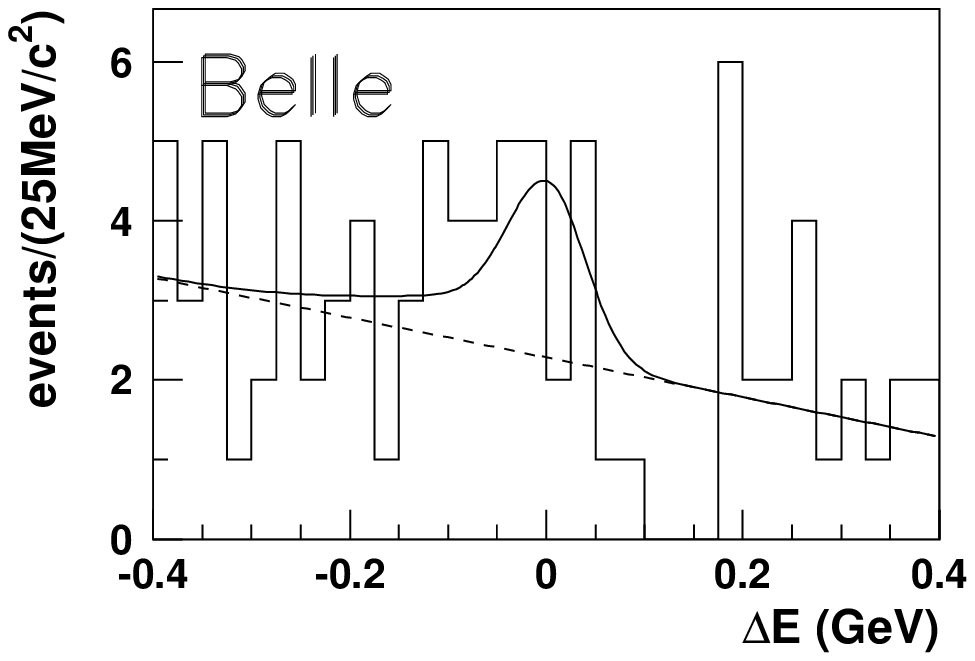}
  \caption{The $\Delta E$ distributions for $B^0 \to \Ktwost^0 \gamma$ (left)
  and $B^+ \to \Ktwost^+ \gamma$ (right) candidates.
  }
  \label{fig:defit-k2st-data}
 \end{center}
\end{figure}

The event selection efficiency is determined from a MC sample that is
calibrated with high statistics control data samples.
% for all final state particles.
Table \ref{tab:systematics} summarizes the sources of systematic error;
a detailed description is given in Ref.~\cite{PL-xsgamma}.
The signal efficiency is $\EFFktwostgammN$ for $B^0 \to \Ktwost^0 \gamma$,
%and $\EFFktwostgammC$ for $B^+ \to \Ktwost^+ \gamma$,
%and $\EFFkstAgammN$ for $B^0 \to K^*(1410)^0 \gamma$,
including the sub-decay branching fractions.
%Then, we obtain
%\begin{eqnarray*}
% {\cal B}(B^0 \to ``\Ktwost^0\mbox{''} \gamma) & = & \BRktwostgammN \\
% {\cal B}(B^+ \to ``\Ktwost^+\mbox{''} \gamma) & = & \BRktwostgammC
%\end{eqnarray*}
%where the quotation mark indicates that no contamination
%from $B \to K^*(1410) \gamma$ or non-resonant decay is assumed.

\begin{table}
 \begin{center}
  \catcode`;=\active \def;{\phantom{0}}
  \caption{Contributions to the systematic error.}
  \begin{tabular}{lccc}
   & $B \to \Ktwost \gamma$ & $B \to K \pi \pi \gamma$ \\ \hline
   photon reconstruction            & 5.3\% & 5.3\%  \\
   charged track reconstruction     & 3.0\% & 4.4\%  \\
   charged kaon selection           & 1.7\% & 1.7\%  \\
   charged pion selection           & 0.6\% & 1.2\%  \\
%   $K_S^0$ reconstruction       & & --- & --- \\
%   $\pi^0$ reconstruction       & & --- & --- \\
   L.R. + $\pi^0/\eta$ veto + vertex & 2.4\% & 5.1\%  \\
%   contamination from $\bsgamma$ process
%                                    & 0.9\% & 1.6\% & ;5.3\% \\
   sub branching ratio uncertainty  & 2.4\%  & --- \\ \hline
   total                        & 7.2\% & 9.1\%
  \end{tabular}
  \label{tab:systematics}
 \end{center}
\end{table}

In order to distinguish the $B \to \Ktwost \gamma$ signal
from $B \to K^*(1410) \gamma$ and non-resonant decays,
we examine the helicity angle distribution for
the signal candidates.
All three modes have different helicity
distributions:
%Using the neutral $B$ decay candidates, we examine
%the invariant $K\pi$ mass distribution and the decay helicity
%angle of the candidates.
%To distinguish the $B \to \Ktwost \gamma$ signal with
%$B \to K^*(1410) \gamma$ or non-resonant decay,
%$\Ktwost$ decay helicity angle ($\coshel$) is useful,
%because the distribution for $\coshel$ is
$\cos^2\thetahel - \cos^4\thetahel$ for $\Ktwost$,
$1 - \cos^2\thetahel$ for $K^*(1410)$ and uniform for
non-resonant decay.
We divide $\coshel$ into 5 bins,
and extract the yield from fits to the $\Mbc$ distribution for each bin
(Fig.~\ref{fig:ycoshel-k2st}).
This distribution clearly favors $B \to \Ktwost \gamma$.
We fit the $\coshel$ distribution
with a function which is artificially parametrized
to avoid a negative yield of $K^*(1410)$ or non-resonant component,
%\[
% f(x) = a \frac{15}{2}(x^2-x^4)
% + b \left( \frac{1}{c^2+1} \frac{3}{2}(1-x^2) + \frac{c^2}{c^2+1} \right),
%\]
and obtain $\YktwostgammNtrue$ events
for the $B \to \Ktwost \gamma$ component.
After subtracting other $\bsgamma$ contributions,
this leads to a $B^0 \to \Ktwost^0 \gamma$ branching
fraction of
\[
 {\cal B}(B^0 \to \Ktwost^0 \gamma) = \BRktwostgammNtrue.
\]
This result agrees with the prediction
from the relativistic form-factor model
of Veseli and Olsson~\cite{PL-VeseliOlsson},
but is much lower than that from
the non-relativistic form-factor model of Ali,
Ohl and Mannel~\cite{PL-AliOhl}.
We also obtain the upper limit
${\cal B}(B^0 \to K^*(1410)^0 \gamma) < \BRULkstAgammN~(90 \% \mbox{~C.L.})$
%for $B \to K^*(1410) \gamma$ signal
by conservatively neglecting the non-resonant component
in the fitting procedure.

The background subtracted $K\pi$ invariant
mass distribution for
$B \to K \pi \gamma$ is obtained by a similar method.
We divide the $M_{K\pi}$ spectrum into
$100 \mbox{~MeV}/c^2$ bins, and extract the signal yield
from the $\Mbc$ distribution for each bin.
In Fig.~\ref{fig:ykxmass-k2st}.
we see a clear enhancement
around $1.4 \mbox{~GeV}/c^2$,
%which indicates the small contribution from non-resonant decay.
which supports the conclusion that the $B \to \Ktwost \gamma$
contribution dominates.

\begin{figure}[!tb]
 \begin{center}
  \begin{minipage}{\MINIPAGEWIDTH}
  \epsfxsize 0.97\textwidth \epsfbox{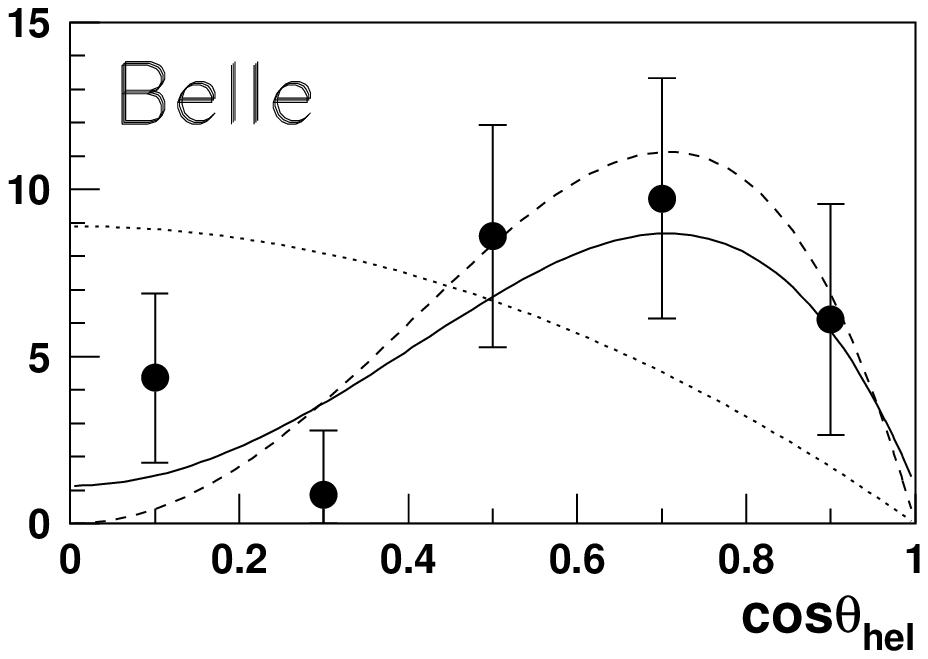}
  \caption{The background subtracted
   $\Ktwost$ helicity angle distribution.
%   Backgrounds are subtracted in every points.
   The solid curve is the fitting result.
   The theoretical curve for $B \to \Ktwost \gamma$
   ($B \to K^*(1410) \gamma$) is shown as
   the dashed (dotted) line.
   }
  \label{fig:ycoshel-k2st}
  \end{minipage}
  \hfil
  \begin{minipage}{\MINIPAGEWIDTH}
  \epsfxsize 0.97\textwidth \epsfbox{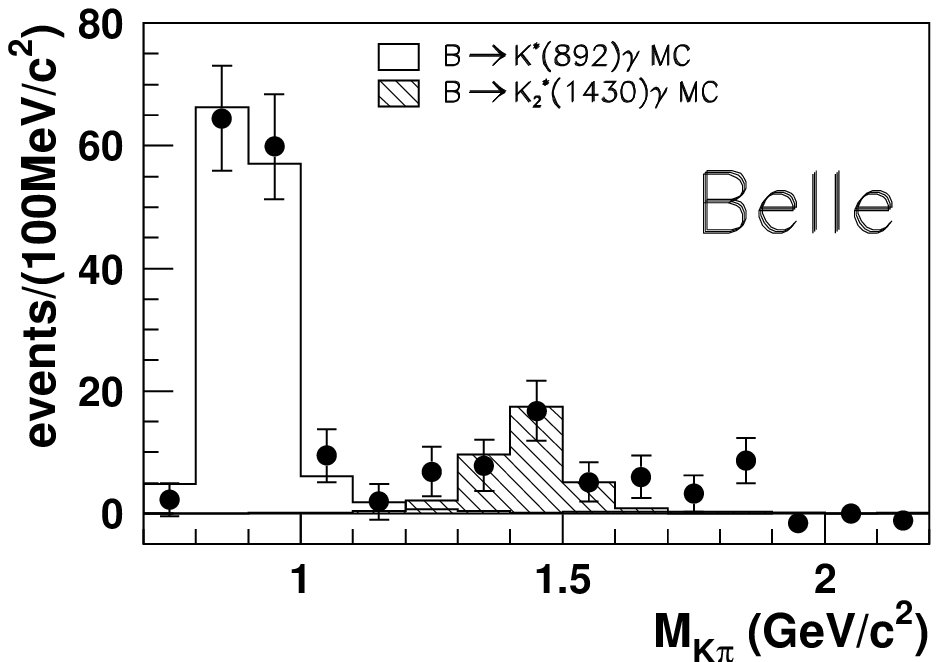}
  \caption{The background subtracted
   $K\pi$ invariant mass distribution.
   }
  \label{fig:ykxmass-k2st}
  \end{minipage}
 \end{center}
\end{figure}

\section{Analysis of $B \to K_X \gamma \to K \pi \pi \gamma$}

The selection criteria used to reconstruct the $B \to K \pi \pi \gamma$
decay
are identical to those used in the analysis of $B \to \Ktwost \gamma$,
unless explicitly stated otherwise.
The $K_X$ candidate is reconstructed from $K^+ \pi^- \pi^+$,
and required to have a mass between 1.0 $\mbox{GeV}/c^2$ and
2.0 $\mbox{GeV}/c^2$.
The three charged tracks are required to form a vertex.
%consistent with a $B$ meson decay vertex.

We select $B \to K_X \gamma \to K^* \pi \gamma$ candidates
(here and throughout this section, $K^*$ denotes $K^*(892)$
for simplicity)
by requiring the invariant mass of $K^+\pi^-$ to be
within $\pm 75 \mbox{~MeV}/c^2$ of the nominal $K^*$ mass.
The resulting $\Mbc$ and $\Delta E$ distributions
are shown in Figs.~\ref{fig:mbfit-kstpi-data} and \ref{fig:defit-kstpi-data},
respectively.
Using the same fitting procedure as is used for the $B \to \Ktwost \gamma$
analysis, we obtain
$\Ykstpigamm$ events from the $\Mbc$ distribution.
This is consistent with the yield obtained from the $\Delta E$ distribution.
These events are dominated by $B \to K^* \pi \gamma$
as a $K^*$ mass peak is clearly seen in the $K\pi$ invariant
mass distribution in Fig.~\ref{fig:ykstmass-kstpi}.
However, other contributions, which can arise either
from $B^+ \to K^+ \rho^0 \gamma$ or
non-resonant $B^+ \to K^+ \pi^- \pi^+ \gamma$,
are not negligible.
We estimate these contributions to be $\YkstpigammBKG$ events
from the region
$1.1 \mbox{~GeV}/c^2 < M_{K\pi} < 1.4 \mbox{~GeV}/c^2$.
An additional contribution from other $\bsgamma$ decay
is estimated to be $\YkstpigammSUBTbsgamma$ events from MC.
After subtracting these non $K^*$ contributions,
we obtain a $B^+ \to K^{*0} \pi^+ \gamma$ yield
of $\YkstpigammSUBTSUBT$ events.

From the $K_X$ invariant mass ($M_{K_X}$)
distribution (Fig.~\ref{fig:ykxmass-kstpi}),
we observe a broad structure below $2.0 \mbox{~GeV}/c^2$
that can be explained, for example, as a sum of two
known resonances around $1.4 \mbox{~GeV}/c^2$
and $1.7 \mbox{~GeV}/c^2$, but cannot be explained by
a single known resonance or phase space decay. We observe no excess above
$2.0 \mbox{~GeV}/c^2$, indicating that the $M_{K_X} < 2.0 \mbox{~GeV}/c^2$
cut does not introduce a significant inefficiency.

To estimate the efficiency of $B \to K^* \pi \gamma$,
we analyze $B \to K_1(1400) \gamma$ and $B \to K^*(1680) \gamma$
MC samples, use the mean of the efficiencies as the central value,
and assign the difference to the systematic error.
As a result, the efficiency becomes $\EFFkstpigamm$
including the other systematic errors in
Table \ref{tab:systematics}.
We determine the $B\to K^* \pi \gamma$ branching fraction,
\[
 {\cal B}(B \to K^* \pi \gamma; M_{K^*\pi}<2.0\mbox{~GeV}/c^2 )
 = \BRkstpigamm.
\]

\begin{figure}[!tb]
 \begin{center}
  \begin{minipage}{\MINIPAGEWIDTH}
  \epsfxsize 0.97\textwidth \epsfbox{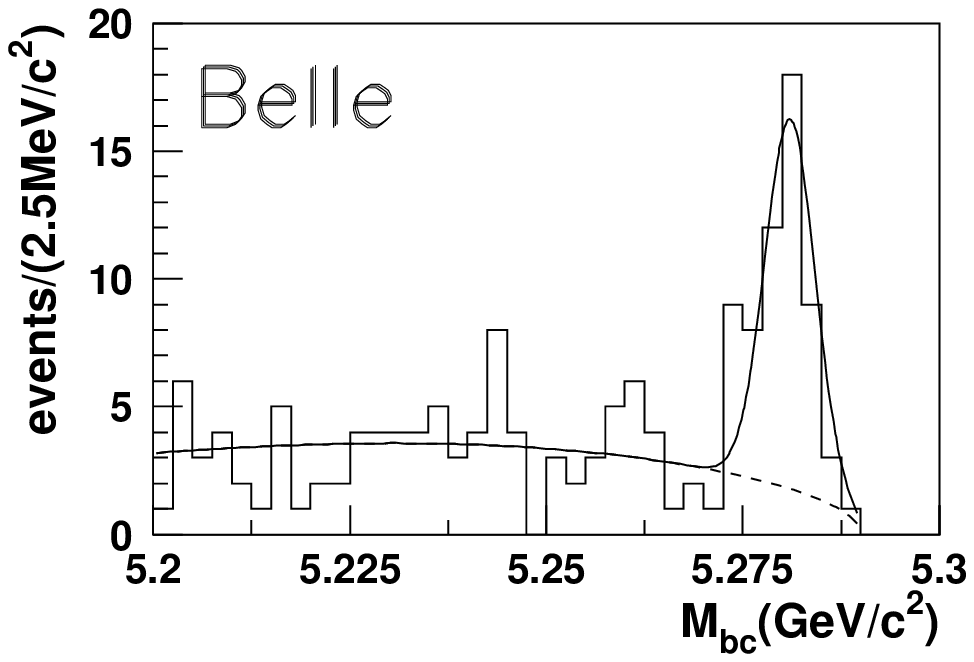}
  \caption{The $\Mbc$ distribution for $B \to K^* \pi \gamma$
   candidates.}
%   The curve is the result of fitting with a single Gaussian
%   (solid line) on top of the background shape (dashed line)
%   determined from the sideband analysis.}
  \label{fig:mbfit-kstpi-data}
  \end{minipage}
  \hfil
  \begin{minipage}{\MINIPAGEWIDTH}
  \epsfxsize 0.97\textwidth \epsfbox{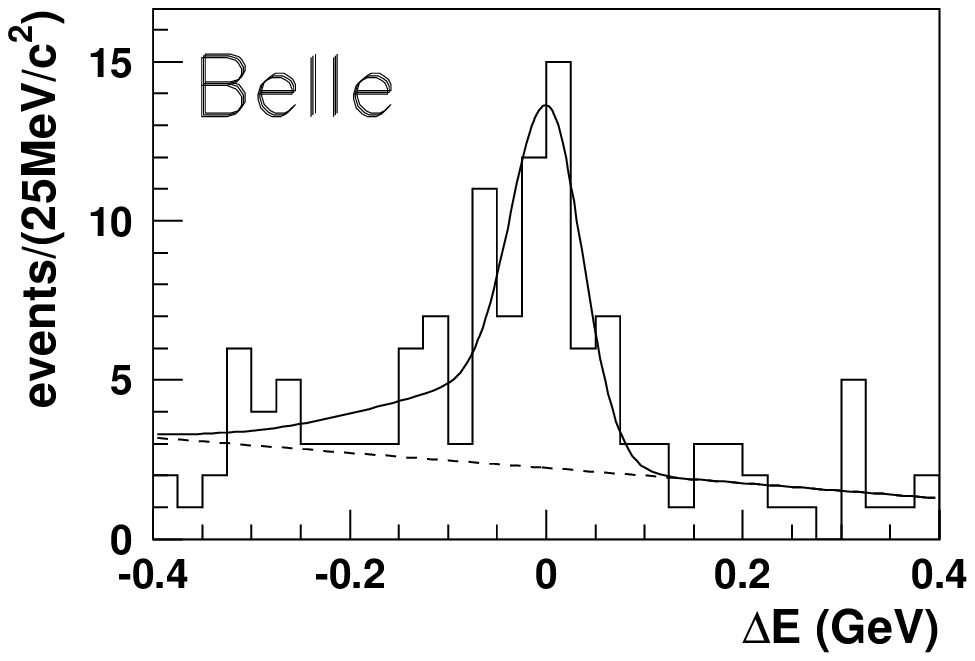}
  \caption{The $\Delta E$ distribution for $B \to K^* \pi \gamma$
   candidates.}
  \label{fig:defit-kstpi-data}
  \end{minipage}
 \end{center}
\end{figure}

\begin{figure}[!tb]
 \begin{center}
  \begin{minipage}{\MINIPAGEWIDTH}
  \epsfxsize 0.97\textwidth \epsfbox{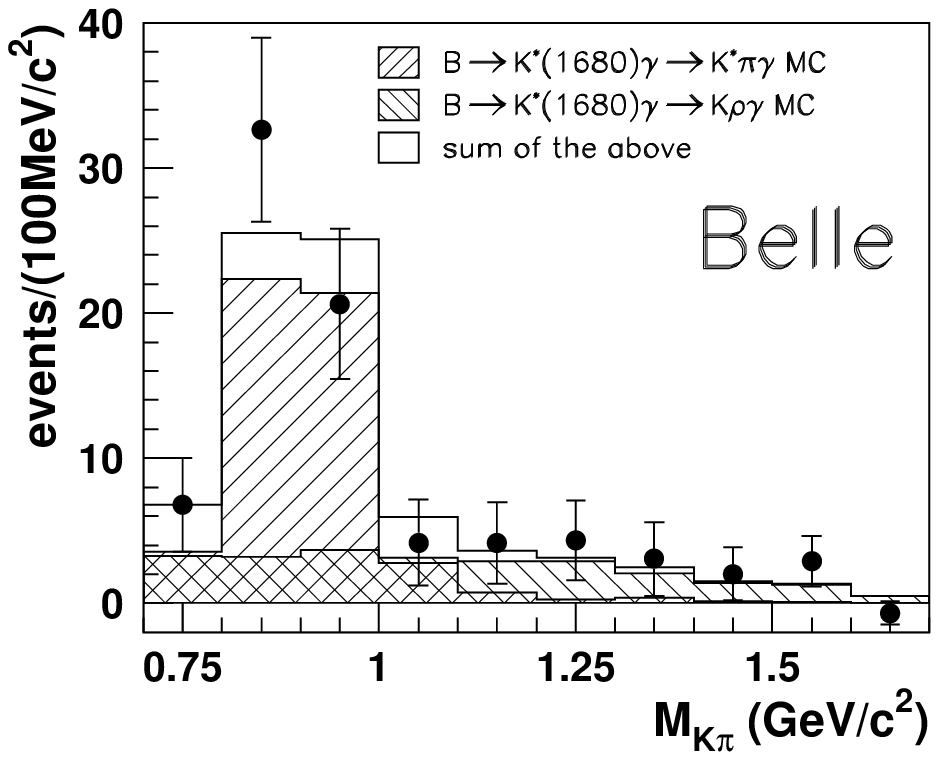}
  \caption{The background subtracted $K^*$ invariant mass distribution
   for the $B \to K^* \pi \gamma$ analysis.
   }
  \label{fig:ykstmass-kstpi}
  \end{minipage}
  \hfil
  \begin{minipage}{\MINIPAGEWIDTH}
  \epsfxsize 0.97\textwidth \epsfbox{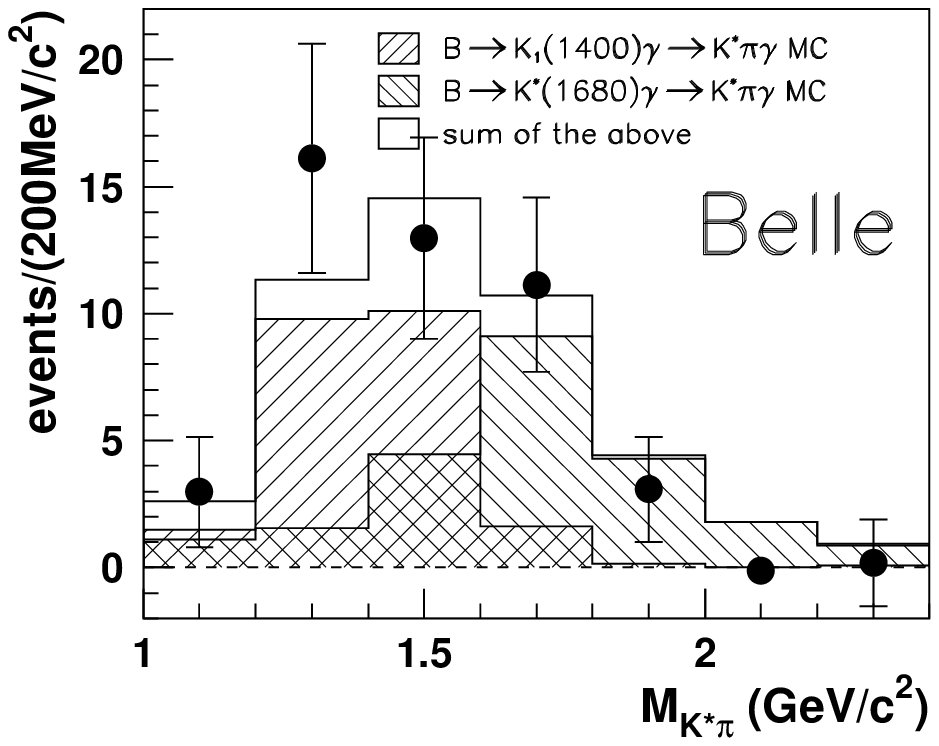}
  \caption{The background subtracted $K\pi\pi$ invariant mass distribution
   for the $B \to K^* \pi \gamma$ analysis.}
  \label{fig:ykxmass-kstpi}
  \end{minipage}
 \end{center}
\end{figure}

There are four known resonances, $K_1(1270)$,
$K_1(1400)$, $K^*(1410)$ and
$\Ktwost$, that can contribute to the signal
around $M_{K_X} = 1.4 \mbox{~GeV}/c^2$. In
the region of $1.2\mbox{~GeV}/c^2 < M_{K_X} < 1.6 \mbox{~GeV}/c^2$,
we obtain $\YkstpigammZ$ events
from the $\Mbc$ distribution.
The $\Ktwost$ contribution is estimated to
be $\NktwostINkstpi$ events from our branching fraction measurement,
assuming $B^0 \to \Ktwost^0 \gamma$
and $B^+ \to \Ktwost^+ \gamma$ have equal branching fractions.
The reconstruction efficiencies are about the same
for $K_1(1400)\gamma$ and $K^*(1410)\gamma$ and
a factor of two lower for $K_1(1270)\gamma$ including
the $K_1(1270)\to K\rho$ and $K_1(1270)\to K_0^*(1430)\pi$
contributions.
%As the
%$K^*(1410)\gamma$ and $K_1(1270)\gamma$ branching fractions from other decay
%channels are poorly constrained as discussed elsewhere in this
%report, we have to assume an equal possibility for
%each of these three resonances.
%In addition, higher resonances may have a tail below $1.6 \mbox{~GeV}/c^2$.
We interpret the signal yield as an upper limit on the weighted sum of the
three resonances,
\begin{eqnarray*}
 \frac{1}{2} {\cal B}(B \to K_1(1270) \gamma) +
 {\cal B}(B \to K_1(1400) \gamma) &+& {\cal B}(B \to K^*(1410) \gamma) \\*
 & < & \ULkstpigammZ~~~(90\% \mbox{~C.L.})
\end{eqnarray*}
This limit on
${\cal B}(B \to K^*(1410)\gamma)$
is more stringent than that obtained from the $\coshel$
distribution of $K\pi\gamma$ decays.
%and the limit of $B \to K_1(1270)\gamma$
%which will be discussed later.

Next, we select $B \to K_X \gamma \to K \rho \gamma$ candidates
by requiring the invariant mass of the $\pi^+\pi^-$ combination to be
within $\pm 250 \mbox{~MeV}/c^2$ of the nominal $\rho$ mass.
To veto $B \to K_X \gamma \to K^* \pi \gamma$ events,
we reject a candidate if the invariant $K^+\pi^-$ mass is
within $\pm 125 \mbox{~MeV}/c^2$ of the nominal $K^*$ mass.
The $\Mbc$ distribution and the $K_X$ invariant mass distribution
are shown in Figs.~\ref{fig:mbfit-krho-data} and \ref{fig:ykxmass-krho},
respectively.
From the $\Mbc$ distribution, we obtain
a signal yield of $\Ykrhogamm$ events.
We subtract the contribution of
$\Ykrhogammbsgamma$ events from
other $\bsgamma$ decays.

The $M_{K_X}$ spectrum of
these events (Fig.~\ref{fig:ykxmass-krho}) shows a large peak
around $1.7 \mbox{~GeV}/c^2$.
One possible explanation is a large
$K^*(1680) \gamma$ component with a small contribution from
$K_1(1270)$, but since
there are quite a few resonances around $1.7 \mbox{~GeV}/c^2$,
a detailed analysis will be required to disentangle the resonant substructure.
%we cannot distinguish them.
The  reconstruction efficiency for
$B \to K \rho \gamma$, which is $M_{K_X}$ dependent, is determined to be
$\EFFkrhogamm$ by assuming a mixture of $K_1(1270)$ and $K^*(1680)$ with a
ratio from the $M_{K_X}$ fit result. So far we
find no signal outside the $\rho$ mass window;
neglecting the non-resonant $K\pi\pi\gamma$ contribution,
we determine the $B\to K \rho \gamma$ branching fraction,
\[
 {\cal B}(B \to K \rho \gamma; M_{K\rho} < 2.0\mbox{~GeV}/c^2 )
 = \BRkrhogamm.
\]

The $K\rho\gamma$ final state in the mass range around $1.3 \mbox{~GeV}/c^2$
is effective for the search of $B \to K_1(1270) \gamma$,
because the $K_1(1270)$ has a large ($(42\pm6)\%$) branching fraction
to $K\rho$.
The second largest contribution is from
$\Ktwost \to K \rho$ ($(8.7\pm0.8)\%$) from which
we expect less than one event. Other contributions are much
smaller.
We search for $B\to K_1(1270) \gamma$ decays
by requiring $|M_{K_X} - M_{K_1(1270)}| < 0.1 \mbox{~GeV}/c^2$,
as shown in Fig.~\ref{fig:mbfit-krho-kx13-data}.
We find $\YkrhogammNSIGBOX$ candidates in the signal box
with a background expectation of $\YkrhogammNBKG$ events.
Using a reconstruction efficiency of $\EFFkoneAgamm$,
we obtain an upper limit of
${\cal B}(B \to K_1(1270) \gamma) < \ULkoneAgammN ~(90\% \mbox{~C.L.})$.
%which is less stringent than the limit obtained in the
%$K^*\pi\gamma$ analysis.

\begin{figure}[!tb]
 \begin{center}
  \begin{minipage}{\MINIPAGEWIDTH}
  \epsfxsize 0.97\textwidth \epsfbox{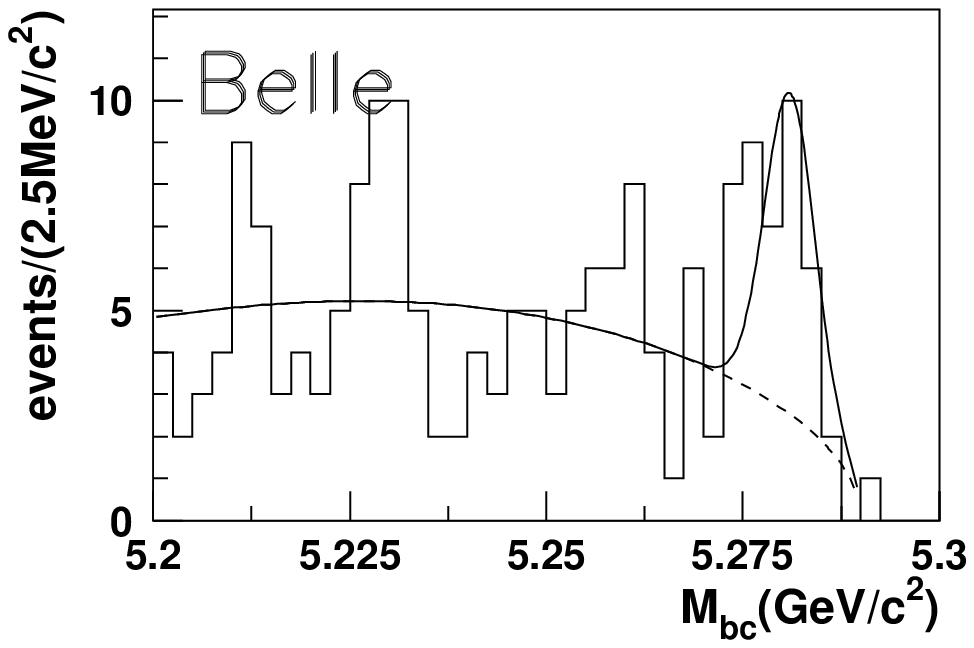}
  \caption{The $\Mbc$ distribution for $B \to K \rho \gamma$
   candidates.}
%   The curve is the result of fitting with a single Gaussian
%   (solid line) on top of the background shape (dashed line)
%   determined from the sideband analysis.}
  \label{fig:mbfit-krho-data}
  \end{minipage}
  \hfil
  \begin{minipage}{\MINIPAGEWIDTH}
  \epsfxsize 0.97\textwidth \epsfbox{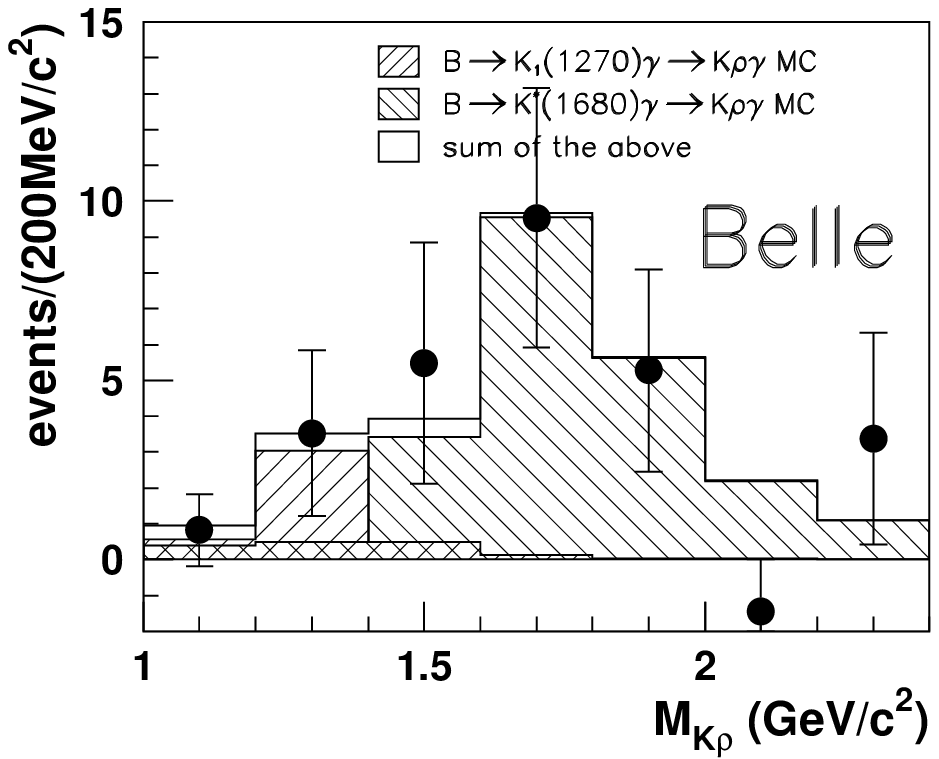}
  \caption{The $K\pi\pi$ invariant mass distribution
   in the $B \to K \rho \gamma$ analysis.
   Background is subtracted in each bin.
   }
  \label{fig:ykxmass-krho}
  \end{minipage}
 \end{center}
\end{figure}

\begin{figure}[!tb]
 \begin{center}
  \epsfxsize 0.47\textwidth \epsfbox{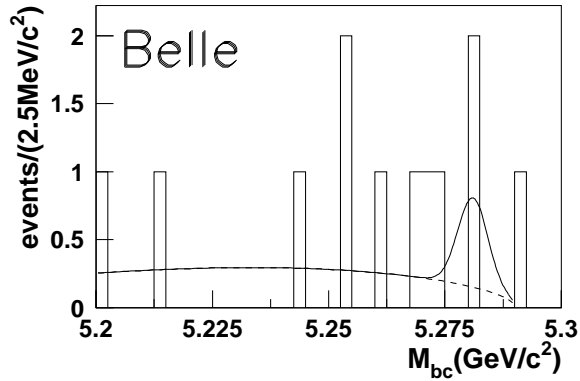}
  \caption{The $\Mbc$ distribution for $B \to K_1(1270) \gamma$
   candidates.}
%   The curve is the result of fitting with a single Gaussian
%   (solid line) on top of the background shape (dashed line)
%   determined from the sideband analysis.}
  \label{fig:mbfit-krho-kx13-data}
%  \hfil
%  \begin{minipage}{\MINIPAGEWIDTH}
%  \epsfxsize 0.97\textwidth \epsfbox{ykstmass-krho.eps}
%  \caption{}
%  \label{fig:ykstmass-krho}
%  \end{minipage}
 \end{center}
\end{figure}

\section{Conclusion}

We have searched for radiative $B$ meson decays into kaonic resonances
that decay into a two-body or three-body final states together with a high
energy photon.  We observe sizable signals
in $B \to \Ktwost \gamma$, $B \to K^*\pi \gamma$
and $B\to K \rho \gamma$ decays and determine the branching
fractions for these channels.  The measured branching fractions
respectively correspond to about 4\%, 17\% and 19\% of the total
$\bsgamma$ branching fraction assuming the SM calculation
~\cite{PL-Misiak} or existing
measurements~\cite{PL-xsgamma,CLEO-xsgamma,ALEPH-xsgamma}
as the denominator.  Adding 15\% from the $K^*(892) \gamma$
branching fractions, these decay modes sum up to about half of the
entire $\bsgamma$ process.

For the $K\pi\gamma$ final state, the $\Ktwost \gamma$ component is separated
from a possible $K^*(1410)\gamma$ or non-resonant contribution using a
helicity angle analysis.
% We also find the branching fraction of $B \to \Ktwost \gamma$
% is about the same for neutral and charged $B$ decays.

For the three-body final states, we observe
$B \to K^*\pi \gamma$ and $B\to K \rho \gamma$
signals separately for the first time;
however, the possible contribution of many kaonic resonances
prevents us from further identification of such resonances
with the current
statistics.  We find no significant signal
for $B\to K_1(1270)\gamma$ decay
in the $K\rho\gamma$ final state.
% and the result
%is consistent with the observation in the $K^*\pi\gamma$ final state as a
%sum of three neighbouring resonances.

\section{Acknowledgement}

%%% \input{acknowledge.tex}
% Please paste this acknowledgement into your latex file.  
%***** Acknowledgments *****
We wish to thank the KEKB accelerator group for the excellent
operation of the KEKB accelerator. We acknowledge support from the
Ministry of Education, Culture, Sports, Science, and Technology of Japan
and the Japan Society for the Promotion of Science; the Australian
Research
Council and the Australian Department of Industry, Science and
Resources; the Department of Science and Technology of India; the BK21
program of the Ministry of Education of Korea and the CHEP SRC
program of the Korea Science and Engineering Foundation; the Polish
State Committee for Scientific Research under contract No.2P03B 17017;
the Ministry of Science and Technology of Russian Federation; the
National Science Council and the Ministry of Education of Taiwan; the
Japan-Taiwan Cooperative Program of the Interchange Association; and
the U.S. Department of Energy.

%%% \input{reference.tex}

%%%


\begin{thebibliography}{99}
 \bibitem{PL-VeseliOlsson}
	 S.Veseli and M.G.Olsson, Phys. Lett. B
	 \textbf{367}, 309 (1996).
 \bibitem{PL-AliOhl}
	 A.Ali, T.Ohl and T.Mannel, Phys. Lett. B
	 \textbf{298}, 195 (1993)
 \bibitem{PRL-CLEO-radb}
	 CLEO Collaboration, T.E.Coan \textit{et al.},
	 Phys. Rev. Let. \textbf{84}, 5283 (2000).
 \bibitem{NIM}
	 Belle Collaboration, K. Abe {\it et al.},
	 KEK Progress Report 2000-4 (2000),
	 to be published in Nucl. Inst. and Meth. A.
 \bibitem{kekb}
	 KEKB B Factory Design Report, KEK Report 95-7 (1995),
	 unpublished; Y. Funakoshi {\it et al.}, Proc. 2000
	 European Particle Accelerator Conference, Vienna (2000).
% \bibitem{bib:belle-tdr}
%	 Belle Collaboration,  Technical Design Report,
%	 KEK Report 95-1, 1995.
% \bibitem{bib:belle-cdc}
%	 H.~Hirano {\it et al.}, KEK Preprint 2000-2. submitted to
%	 Nucl. Instr. Meth. A;
%	 M.~Akatsu {\it et al.}, DPNU-00-06, submitted to Nucl. Instr. Meth. A.
% \bibitem{bib:belle-svd}
%	 G.~Alimonti {\it et al.}, KEK preprint 2000-34.
% \bibitem{bib:belle-acc}
%	 T.~Iijima {\it et al.},
%	 ``Aerogel Cherenkov counter for the Belle detector'',
%	 in the Proceedings of the 7th International Conference
%	 on Instrumentation for Colliding Beam Physics,
%	 Hamamatsu, Japan, Nov. 15--19, 1999.
% \bibitem{bib:belle-tof}
%	 H.~Kichimi {\it et al.}, submitted to Nucl. Instr. Meth. A.
% \bibitem{bib:belle-csi}
%	 H.~Ikeda et al., Nucl. Instr. Meth. {\bf A441}, 401 (2000).
 \bibitem{PL-xsgamma}
	 Belle Collaboration, K.Abe {\it et al.}, Phys. Lett. B
	 \textbf{511}, 151 (2001).
% \bibitem{fw}
%	 G. Fox and S. Wolfram, Phys. Rev. Lett {\bf 41}, 1581 (1978).
 \bibitem{argus}
	 ARGUS Collaboration,
	 H.~Albrecht {\it et al.}, Phys. Lett. B {\bf 241}, 278 (1990).
 \bibitem{bib:pdg}
	 Particle Data Group, D.E. Groom {\it et al.},
	 Eur. Phys. J. {\bf C15}, 1 (2000).
 \bibitem{PL-Misiak}
	 K.Chetyrkin, M.Misiak, M.M\"unz,  Phys. Lett. B
	 \textbf{400}, 206 (1997);
	 Erratum ibid. B \textbf{425}, 414 (1998)
 \bibitem{CLEO-xsgamma}
	 CLEO Collaboration, M.Alam {\it et al.}, Phys. Rev. Lett.
	 {\bf 74}, 2885 (1995).
 \bibitem{ALEPH-xsgamma}
	 ALEPH Collaboration, R.Barate {\it et al.}, Phys. Lett. B
	 \textbf{429}, 196 (1998)
\end{thebibliography}
\end{document}